\documentstyle[preprint,aps,floats,epsfig]{revtex}

\def\book#1[[#2]]{{\it#1\/} (#2).}
\def\am#1 #2 #3.{{\it Ann.\ Math.\ \bf#1} #2 (#3).}
\def\apj#1 #2 #3.{{\it Astrophys.\ J.\ \bf#1} #2 (#3).}
\def\atmp#1 #2 #3.{{\it Adv.\ Theor.\ Math.\ Phys.\ \bf#1} #2 (#3).}
\def\cmp#1 #2 #3.{{\it Commun.\ Math.\ Phys.\ \bf#1} #2 (#3).}
\def\comnpp#1 #2 #3.{{\it Comm.\ Nucl.\ Part.\ Phys.\  \bf#1} #2 (#3).}
\def\cqg#1 #2 #3.{{\it Class.\ Quant.\ Grav.\ \bf#1} #2 (#3).}
\def\epl#1 #2 #3.{{\it Europhys.\ Lett.\ \bf#1} #2 (#3).}
\def\grg#1 #2 #3.{{\it Gen.\ Rel.\ Grav.\ \bf#1} #2 (#3).}
\def\jmp#1 #2 #3.{{\it J.\ Math.\ Phys.\ \bf#1} #2 (#3).}
\def\ijmpd#1 #2 #3.{{\it Int.\ J.\ Mod.\ Phys.\ \bf D#1} #2 (#3).}
\def\mpla#1 #2 #3.{{\it Mod.\ Phys.\ Lett.\ \rm A\bf#1} #2 (#3).}
\def\ncim#1 #2 #3.{{\it Nuovo Cim.\ \bf#1\/} #2 (#3).}
\def\npb#1 #2 #3.{{\it Nucl.\ Phys.\ \rm B\bf#1} #2 (#3).}
\def\phrep#1 #2 #3.{{\it Phys.\ Rep.\ \bf#1\/} #2 (#3).}
\def\pla#1 #2 #3.{{\it Phys.\ Lett.\ \bf#1\/}A #2 (#3).}
\def\plb#1 #2 #3.{{\it Phys.\ Lett.\ \bf#1\/}B #2 (#3).}
\def\pr#1 #2 #3.{{\it Phys.\ Rev.\ \bf#1} #2 (#3).}
\def\prd#1 #2 #3.{{\it Phys.\ Rev.\ \rm D\bf#1} #2 (#3).}
\def\prl#1 #2 #3.{{\it Phys.\ Rev.\ Lett.\ \bf#1} #2 (#3).}
\def\prs#1 #2 #3.{{\it Proc.\ Roy.\ Soc.\ Lond.\ A.\ \bf#1} #2 (#3).}

\def\half{\textstyle{1\over2}}

\font\cat=cmr7
\def\E{\text{\cat E}}
\def\T{\text{\cat T}}
\def\U{${\cal U}$}
\newcommand{\be}{\begin{equation}}
\newcommand{\ee}{\end{equation}}
\newcommand{\bea}{\begin{eqnarray}}
\newcommand{\eea}{\end{eqnarray}}
\newcommand{\bml}{\begin{mathletters}}
\newcommand{\eml}{\end{mathletters}}

\begin{document}
\preprint{DTP/00/59, hep-th/0007177}
\draft
\tighten

\title{General brane cosmologies and their global spacetime structure}
\author{Peter Bowcock\footnote{E-mail 
address: \texttt{Peter.Bowcock@durham.ac.uk}}, Christos 
Charmousis\footnote{E-mail address: \texttt{Christos.Charmousis@durham.ac.uk}}, 
Ruth Gregory.\footnote{E-mail address: \texttt{R.A.W.Gregory@durham.ac.uk}}}
\address{Centre for Particle Theory, 
         Durham University, South Road, Durham, DH1 3LE, U.K.}
\date{\today}
\setlength{\footnotesep}{0.5\footnotesep}
\maketitle

\begin{abstract}
Starting from a completely general standpoint, we find the most general
brane-Universe solutions for a three-brane in a five dimensional
spacetime. The brane can border regions of spacetime with or without
a cosmological constant. Making no assumptions other than the usual
cosmological symmetries of the metric, we prove that the equations of motion
form an integrable system, and find the exact solution. The cosmology is
indeed a boundary of a (class II) Schwarzschild-AdS spacetime, or a
Minkowski (class I) spacetime. We analyse the various cosmological
trajectories focusing particularly on those bordering vacuum spacetimes.
We find, not surprisingly, that not all cosmologies are compatible with
an asymptotically flat spacetime branch. We comment on the role of the 
radion in this picture.
\
\end{abstract}

\pacs{PACS numbers: 04.50.+h, 11.25.Mj, 98.80.-k \hfill hep-th/0007177}
\section{Introduction}

The idea that our four-dimensional universe might in fact, upon closer 
inspection, be higher dimensional has always been compelling, although
it has been conventionally supposed that the extra dimensions are small
and compact. An almost diametrically opposed view, wherein the extra
dimensions are large and noncompact \cite{RS} has gained ground recently, 
in which our universe is a `defect' (see \cite{RSA} for early work) in 
an anti-de Sitter bulk -- the negative cosmological constant providing
an effective or exotic compactification \cite{EXO} of the spacetime. In
most models, the universe is taken to be a domain wall, (i.e.\ 
of codimension one -- although models of higher codimension have been 
considered \cite{HCD}) since this is most naturally supported by heterotic 
string theory \cite{HW,LOSW,LOW,HR}. The original models of Randall and Sundrum
\cite{RS} focussed on the main features of having either one or two
domain wall universes at orbifold fixed points of a single extra
dimension; in the case of the two-wall model the second wall had negative
energy and tension. The compelling feature of these brane-world models
is that gravity has a four-dimensional
character on the universe-domain-wall with short-range five-dimensional
corrections coming from Kaluza-Klein (KK) modes. Indeed, it was 
demonstrated explicitly by Garriga and Tanaka \cite{GT} that linearized
gravity on the brane was indeed Einstein gravity for a single wall
universe (see also \cite{GKR}). The key feature of the 
Randall Sundrum (RS) (and indeed many other)
models is that the induced metric on the wall (or walls) is flat
Minkowski spacetime. In general, a domain wall spacetime is not static,
but has a de-Sitter expansion along its spatial directions, however in the
presence of a bulk cosmological constant, the effect of the wall energy
and tension is neutralized, and a static solution is allowed. 
However, our universe is not Minkowski spacetime, nor is it a localized
linearized perturbation thereof. Therefore, if brane universe models are to be
relevant, cosmological brane solutions are the obvious next step.

A general homogeneous and isotropic brane cosmology is simply  a
wall with energy, \E, and tension, \T, which are no longer fixed by
the special {\it domain} wall relation $\E={\T}$. For example, we might
wish to set
\be\label{edef}
\E = \E_0 + \rho \;\;\;\; ; \;\;\;\;\;\; \T = \T_0 - p
\ee
and then take $\rho$ and $p$ to be our `cosmological' energy density and
pressure respectively. This is what is conventionally done, with $\rho$
and $p$ obeying an equation of state (rather than \E\ and \T), although it
should be noted that when $\rho$ and \E\ are of a similar order of
magnitude, there is no reason to suppose that the equation of state will
continue to hold with this choice of $\rho$ and $p$, which become rather
arbitrary at that point.

Much of the work on brane-cosmological models to date has taken a
`brane-based' approach, notably the work of Binetruy et.\ al.\
\cite{BDL,BDEL} (see also \cite{COS}), in which the RS spacetime 
is generalized to allow for time-dependent cosmological expansion.
However, Binetruy et.\ al.\ were concerned primarily with
deriving the four-dimensional brane-cosmological equations, and
for simplicity when finding explicit solutions,
chose a Gaussian Normal gauge in which $g_{nn}$, the
component of the metric normal to the wall, was set to unity. Therefore,
while the brane cosmology was readily apparent in their approach, the
bulk spacetime structure was less transparent, with coordinate singularities
indicating the breakdown of the rather restrictive GN gauge. 
A more general brane-based approach is epitomized by the work of 
Shiromizu et.\ al.\ \cite{SMS} (see also Maartens, \cite{M}, for an
excellent brane-based description of bulk effects) in which the 
Gauss-Codazzi formalism is used to obtain a pure-brane `Einstein'
equation, in which the bulk geometry is encoded in a single tensor
${\cal E}_{\mu\nu}$. While such a brane-based approach has the attraction
of being a simple generalization of the standard four-dimensional
FRW equations, the effect of the bulk on the brane is somewhat less
transparent, with the tensor ${\cal E}_{\mu\nu}$ hiding a multitude of sins. 

An alternate approach for deriving cosmological solutions was taken by
Ida \cite{Ida}, who instead considered a `bulk-based' point of view,
in which the most general static AdS solution with the appropriate symmetries:
\be\label{schads}
ds^2 = h(r) dt^2 - h^{-1}(r) dr^2 - r^2 \left [ {d\chi^2\over1-\kappa\chi^2}
+\chi^2 d\Omega_{I\!I}^2 \right]
\ee
(where $ h(r) = \kappa - {\mu\over r^2} + k^2 r^2 $)
is taken, and the brane becomes an arbitrary boundary of two versions of
this spacetime, with possibly differing masses on each side. 
The Israel-Gauss-Codazzi equations then give the energy and tension of
the boundary as a function of its trajectory. The beauty of this
approach is that it is quite general and does not rely on any $Z_2$ symmetry
around the wall itself; it is also straightforward to derive
brane-cosmological evolution equations -- the cosmological solution then
becomes a wall moving in AdS spacetime; the only disadvantage is that it
is perhaps a little more abstract than the brane-based approach. 
A coordinate transformation relating the solutions of \cite{BDL,BDEL} to 
Ida's solutions was found by by Mukohyama et.\ al.\ \cite{MSM}.
Ida's work in fact generalizes the work of Kraus \cite{PK} (see also
\cite{CRP} for related string theoretic work, and
\cite{Gub}, for an AdS/CFT perspective on the issue), who derived
the most general $Z_2$ symmetric {\it domain} wall solutions by taking
slicings of (\ref{schads}), and so can apply to models with different
cosmological constants such as the Lykken-Randall (LR) model, 
\cite{LR}, for example. 

What we aim to do in this paper is to bridge the gap between the `brane-based' 
approach, where the brane represents a {\it fixed} boundary in some 
time-evolving spacetime, and the `bulk-based' approach, where the 
wall (or spacetime boundary) follows some timelike 
trajectory in a {\it static} bulk spacetime. We start off by considering the
most general brane-based formalism,  following the approach of Ipser and 
Sikivie \cite{IS}, finding the general wall solutions for a fixed wall 
of constant spatial curvature embedded in a bulk of constant curvature. 
We then demonstrate how a {\it fixed wall} embedded in 
a {\it non-static spacetime} is strictly equivalent to a 
{\it moving wall} embedded in a {\it static spacetime} thereby establishing 
in full generality the 
equivalence between the two approaches. This is true  for a wall of completely
arbitrary equation of state separating spacetimes which may even have a 
different cosmological constant. Having shown this equivalence we then proceed
to find and study the most general wall trajectories, i.e.\
cosmological evolution equations, in a static AdS 
or even flat background. The reason for studying such a general
set-up is that some more recent generalizations of the RS scenario have
involved not only non $Z_2$ symmetric walls, such as the 
`Millenium Model' of Kogan et.\ al.\ \cite{KMPR} consisting of two positive
tension walls at orbifold fixed points with an additional negative tension
wall freely moving inbetween, but also patching together of spacetimes
with different cosmological constant, such as the GRS model \cite{GRS}, 
in which the central $Z_2$ symmetric wall is flanked by two negative tension
branes with flat space in the exterior. The curious feature of these
apparently contrived models is that gravity not only changes nature at 
short but also at ultra-large scales, becoming weaker for the Millenium
model, and five-dimensional for the GRS model. (In fact gravity for
the GRS model has several peculiarities, \cite{Probs,GRSPRZ,CEHT}, 
which may be ameliorated 
by a hybrid double-wall variant of the Millenium model, \cite{KMPR2},
in which the negative tension wall of the Millenium model is replaced by
two negative tension walls with a slice of flat space inbetween.)

While we might expect that cosmology in such models is potentially delicate,
the beauty of the geometrical approach is that the search
for a cosmology becomes a local question of finding a suitable trajectory
for the spacetime boundaries, thus the presence of extra walls is 
irrelevant - unless these walls collide. We therefore do not consider
issues such as radion stabilization \cite{COSRAD}, or the like, simply
examining the possible trajectories (and hence cosmological solutions)
of the brane universes. 

In the next section we set up our formalism and then find the general 
solution to the Einstein equations for a constant spatial curvature wall in
a constant curvature spacetime, generalizing Taub's solutions \cite{taub} 
to allow for the presence of a cosmological constant in the bulk. We then show 
how from this brane-based approach we can cross over to the bulk-based 
approach for a wall of arbitrary trajectory evolving in a static spacetime.
In section III we establish the most general cosmological evolution equations
for the different classes of spacetime solutions. In the following section we 
analyze some specific cases of interest in cosmology finding the 
wall trajectories analytically. We make some concluding remarks in Section V.

\section{General wall spacetimes}

In this section we derive the general spacetime of a brane universe,
and the equations of motion it must obey. As per usual, we
shall be modelling our four-dimensional Universe, \U, to be an 
infinitesimally thin wall type defect of constant spatial
curvature embedded in a five-dimensional spacetime. In other words,
we are looking for a spacetime with planar (or spherical/hyperboloidal)
symmetry in three of its spatial directions, which has one (or more) 
hypersurfaces on which the spacetime curvature has a distributional
singularity corresponding to a $\delta$-function source for the wall
energy-momentum. The most general metric admitting this symmetry can
be written in the form
\be\label{bulk1}
ds^2=e^{2\nu(\texttt{t,z})}(B(\texttt{t,z}))^{-2/3}
(d{\texttt t}^2-d{\texttt z}^2)-B^{2/3} 
\left [ {d\chi^2\over1-\kappa\chi^2}+\chi^2 d\Omega_{I\!I}^2 \right]
\ee
where $\kappa = 0, \pm 1$ represents the spatial curvature of the
3-spatial sections, and $\nu, B$ will satisfy the bulk Einstein
equations (with or without a cosmological constant), as well as
appropriate jump conditions at the wall which we will choose to set
at ${\texttt z}=0$.

We start by summarizing the jump conditions for later use. As is conventional,
we denote the normal to \U\ by $n_a$, in terms of which the first
fundamental form, $h_{ab}$, of the wall is given by
\be\label{firstff}
h_{ab}=g_{ab}+n_a n_b
\ee
which is simply the projection of the bulk metric on \U.
The second fundamental form, or extrinsic curvature of \U\ is 
\be\label{secondff}
K_{ab}=h_{(a}^c h_{b)}^d \nabla_c n_d
\ee
The intrinsic and extrinsic geometry of \U\ are related 
to the bulk curvature via the Gauss Codazzi equations
\bml\label{gauscod}\bea
R_{acbd} n^c n^d = K_{ac} K^c_b - {\cal L}_n K_{ab} &=&
R^{(4)}_{ab}  - R_{cd}h_a^c h_b^d - K_{ac}K_b^c + K K_{ab} \label{Gauss} \\
h^{bc} \nabla_b K_{ca} - h_a^b \nabla_b K &=& R^{(4)}_{bc} 
h^b_a n^c  \label{Codazzi}\\
R^{(4)} - K_{ab} K^{ab} + K^2 &=& 2G_{ab} n^a n^b\label{gauss}
\eea\eml
where $R_{ab}^{(4)}$ is the intrinsic Ricci tensor obtained from $h_{ab}$,
and ${\cal L}_n$ is the Lie derivative with respect to $n_a$.

Since the wall's energy-momentum tensor $T_{ab}^{w}$ is a distributional
source, we can now integrate (\ref{Gauss}) across the wall, and defining
\be
S_{ab}=\int T_{ab}^w \, dl
\ee
we immediately arrive at the Israel junction conditions \cite{Israel}
\be
\label{junction}
\Delta K_{ab} = [S_{ab}-\frac{1}{3}S h_{ab}],
\ee
where $\Delta K_{ab} = K^+_{ab} - K^-_{ab}$ is the jump in the extrinsic
curvature, and we have set $8\pi G_5=1$.

Furthermore taking the sum and difference of (\ref{gauss}) and 
(\ref{Codazzi}) respectively and using (\ref{junction}) we have,
\bml \label{G-C} \bea
R^{(4)} - {\overline K_{ab}} {\overline K^{ab}} + {\overline K}^2 
&=& {1\over4}(S_{ab}S^{ab} -{1 \over 3}S^2) +2 {\overline G_{nn}}
\label{g1}\\
{\overline K_{ab}}S^{ab}  &=& 2 \Delta G_{nn} \label{g2}\\
\nabla^{(4)}_bS^{bc} &=& 0 \label{c1}\\
\nabla^{(4)}_b{\overline K^b_a}-\nabla^{(4)}_a {\overline K} &=& 0\label{c2}
\eea
\eml
where ${\overline Q} = (Q^+ + Q^-)/2$ stands for the mean of a quantity 
across the wall. We should stress that (\ref{G-C}) are in fact 
integrability conditions and hence 
will result from the Einstein and junction conditions (\ref{junction}). 

Since we are looking for cosmological solutions, we shall assume that 
our brane Universe is made of homogeneous and isotropic matter, hence
\be
\label{fluid}
S_{ab}=\E u^a u^b+\T(h^{ab}-u^a u^b)
\ee
where $\E$ is the surface energy density and $\T$ the 
tension of the brane 
\U. The timelike vector $u^a$ is the 5-velocity of an observer
comoving with the brane Universe.  Obviously if $\E=\T$ we have a 
{\it domain} wall, and if $\T=0$, a dust wall. 

We begin by finding the most general bulk solution before examining these
boundary junction conditions.
The bulk Einstein equations in this case are simply,
\be
\label{Einstein}
R_{ab}=-{2\over3}\Lambda g_{ab}
\ee
where $\Lambda$ is the cosmological constant. For the metric (\ref{bulk1})
these give the following system of partial differential equations,
\bml
\label{einstein}
\bea
B_{,\texttt{tt}}-B_{,\texttt{zz}} &=& \left (2\Lambda B^{1/3} 
- 6 \kappa B^{-1/3} \right ) e^{2\nu}\label{wave1}\\
\nu_{,\texttt{tt}}-\nu_{,\texttt{zz}} &=& \left ( \frac{\Lambda}{3} B^{-2/3} 
+ \kappa B^{-4/3} \right ) e^{2\nu} \label{wave2}\\
\nu_{,{\texttt z}}B_{,{\texttt t}}+\nu_{,{\texttt t}}B_{,{\texttt z}} 
&=& B_{,\texttt{tz}}\label{int1}\\
2\nu_{,{\texttt z}}B_{,{\texttt z}}+2\nu_{,{\texttt t}}B_{,{\texttt t}} &=& 
B_{,\texttt{tt}}+B_{,\texttt{zz}} \label{int2}
\eea
\eml
The particular way of writing the metric (\ref{bulk1}) shows 
the Liouville like 
character of the system of equations, where (\ref{wave1}), (\ref{wave2}) 
are in the presence of $\Lambda$, non-homogeneous coupled wave 
equations and (\ref{int1}) and (\ref{int2}) will turn out to be 
integrability conditions. This is not too surprising since we 
are effectively studying a 1+1 gravity problem by considering the 
Kaluza-Klein reduction of the constant curvature spacelike dimensions. 

It proves easiest to rewrite (\ref{einstein}) in light-cone coordinates,
\be 
\label{light}
u=\frac{{\texttt t}-{\texttt z}}{2},\qquad v=\frac{{\texttt t}+{\texttt z}}{2}
\ee
in which 
\bml \label{einstein2} \bea
B_{,uv} &=& \left (2\Lambda B^{1/3}  - 6\kappa B^{-1/3} \right )
e^{2\nu}\label{wave11}\\
\nu_{,uv} &=& \left ( \frac{\Lambda}{3} B^{-2/3} + \kappa B^{-4/3} \right)
e^{2\nu}\label{wave12}\\
B_{,u} \left[ln(B_{,u})\right]_{,u} &=& 2\nu_{,u} B_{,u} \label{int11}\\
B_{,v} \left[ln(B_{,v})\right]_{,v} &=& 2\nu_{,v} B_{,v} \label{int12}
\eea \eml

First note that (\ref{int11},\ref{int12}) can be directly integrated
to give
\be
e^{2\nu} = V'(v) B_{,u} = U'(u) B_{,v}
\ee
where $V'(v)$, $U'(u)$ are arbitrary nonzero functions of $u,v$ for generic
$\Lambda,\kappa$. (If $\Lambda=\kappa=0$, then it is possible for
one of $U'(u)$ or $V'(v)$ to vanish -- see below.) It is straightforward
to see then that $B$ and $\nu$ have the form
(where we remind the reader that {\it primes} denote ordinary 
differentiation with respect to the unique variable of the function)
\be
\label{sol1}
B=B(U(u) + V(v)),\qquad e^{2\nu}=B' U' V'
\ee
which reduces the remaining PDE (\ref{wave11}) to the ODE 
\bea \label{sol2}
B''- \left ( 2\Lambda B^{1/3} - 6 \kappa B^{-1/3} \right ) B' &=& 0
\nonumber \\ 
\Rightarrow \qquad B'-{3 \over 2} \Lambda B^{4/3} + 9 \kappa B^{2/3} &=& 9\mu
\eea
where $\mu$ is an integration constant.
This last relation can be integrated to give 
\be
\label{sol3}
{1\over\sqrt{ 9 \kappa^2 - 3\mu\Lambda }} \left [
|r_-| \tan^{-1} {r\over |r_-|} - r_+ \coth^{-1} {r\over r_+} \right]
= \left ( U+V \right )
\ee
where $r=B^{1/3}$, $c$ is a constant of integration, and
\be
r^2_\pm = {3\kappa\over\Lambda} \pm \sqrt{ {9\kappa^2\over \Lambda^2} - 
{6\mu\over \Lambda}} .
\ee
Hence the general solution to (\ref{Einstein}) is given by,
\be
\label{solution}
ds^2=B'U'V'B^{-2/3}(d{\texttt t}^2-d{\texttt z}^2)-B^{2/3}dx_{I\!I\!I}^2
\ee
where $B=B(U+V)$ satisfies (\ref{sol2}) or (\ref{sol3}), $U(u)$ and $V(v)$ 
are arbitrary functions and $dx_{I\!I\!I}^2$ stands for the three dimensional 
constant curvature metric written out explicitly in (\ref{bulk1}).  

When $\Lambda = \kappa = 0$, (\ref{sol2}) gives (wlog) $B=U+V$, and
there are now in fact two allowed classes of solution, in the terminology 
of Taub \cite{taub}; the above, (\ref{solution}), being a class II solution.
The class I solutions are distinguished by having $U'=0$ or $V'=0$, but 
not both, in which case the general metrics are
\bml\bea
\label{class1}
ds^2=\frac{U' H(v)}{U^{2/3}}(d{\texttt t}^2-d{\texttt z}^2)-U^{2/3}d{\bf x}^2\\
ds^2=\frac{V' K(u)}{V^{2/3}}(d{\texttt t}^2-d{\texttt z}^2)-V^{2/3}d{\bf x}^2
\eea\eml
for $U'\neq 0$ and $V'\neq 0$ respectively. In fact these class I solutions
are flat, as we will see in the next section. In summary Einstein's 
equations (\ref{einstein}) admit two distinct classes of solutions I,
(\ref{class1}), and II, (\ref{solution}), which depend on two arbitrary 
functions.

Having derived the general bulk solutions, now let us examine
the constraints, or boundary conditions at ${\texttt z}=0$ imposed by the wall.
For what follows we shall assume $Z_2$ symmetry in the ${\texttt z}$-direction 
to make notation easier and results more transparent.
We  shall however be dropping this assumption when actually looking 
for general cosmological solutions in the next section.

Reflection symmetry about ${\texttt z}=0$ permits us to consider only 
${\texttt z}>0$ in  (\ref{solution}) since the metric is an even function 
with respect to ${\texttt z}$. Extrinsic curvature components 
are thus odd functions in ${\texttt z}$ and hence,
$$
\Delta K_{ab}=2K_{ab}^+, \qquad \overline{K_{ab}}=0
$$

In the coordinate system  adapted to the wall the intrinsic and extrinsic 
geometric quantities are greatly simplified, and take the form 
\bml
\label{z=0}
\bea
\E &=& -2e^{-\nu}B^{-2/3} \partial_z B =
-e^{-\nu}\frac{B'(V'-U')}{B^{2/3}}\label{jun1}\\
({2 \over 3}\E-\T) &=& - 2\partial_z \left ( B^{1/3} e^{-\nu} \right ) = 
B^{1/3}e^{-\nu}\left[{1 \over 2}
\left( \frac{V''}{V'}-\frac{U''}{U'}\right)+\frac{1}{2B}(V'-U')(\Lambda
B^{4/3} -6\mu) 
\right]\label{jun2}
\eea
\eml
Hence (\ref{z=0}) tell us that given two arbitrary functions $U$ and $V$ 
we can determine $\E$ and $\T$. We now show that only one of these functions 
is physical. 

As a general rule any coordinate transformation in the bulk will 
{\it shift} our boundary making our wall evolving in time. However 
we can make a class of coordinate transformations in the bulk which 
leave our wall boundary conditions fixed, namely
\be
\label{conformal}
u\longrightarrow f(u), \qquad v\longrightarrow f(v)
\ee
i.e.\ the rescaling of the light-cone coordinates by some arbitrary function. 
This boundary invariance is a result of the two-dimensional conformal 
symmetry on the $\texttt{t-z}$ plane. Now we choose a particular coordinate 
transformation, $f\equiv V$, to fix this non-physical gauge freedom, thereby
effectively setting $V\equiv v$.
Hence there is only one remaining arbitrary function or physical
degree of freedom, $U(u)$, which
completely determines \E\ and \T\ for the wall.

There are now  two equivalent ways, as it turns out, to look for 
cosmological solutions. We either, in the spirit of \cite{BDEL}, 
consider a fixed boundary in the general spacetime (\ref{solution}),
using the junction conditions (\ref{z=0}), 
or we make coordinate transformations simplifying the bulk 
and study the {\it moving} wall trajectories in that case.

In fact, it turns out that the general class II solution (\ref{solution}) 
is a static spacetime.  Setting
\be \label{coord1}
r = B^{1/3} \;\; ; \;\;\;\;\;\; t = 3(v-U) 
\ee
(\ref{solution}) becomes,
\be \label{staticm}
ds^2= h(r) dt^2 - {dr^2 \over h(r)} - r^2 dx_{I\!I\!I}^2
\ee
where $h(r)$ has the familiar form
\be\label{hrdef}
h(r) = - {B'B^{-2/3}\over 9} = \kappa - {\Lambda\over6} r^2 - {\mu\over r^2}
\ee
recognised as the generalised ``Schwarzschild-AdS'' solutions. 

Note however that in making our background static our wall is no longer 
fixed, since the trajectory $\texttt z = 0$, or $u=v$, gives a nontrivial 
relationship between $r$ and $t$: $r = R(t)$ from (\ref{coord1}).
The reader should note here, that time dependence of the metric 
(\ref{solution}) in the bulk, portrayed by arbitrary function $U(u)$, 
and static boundary 
(\ref{z=0}) has been transformed into time dependence for the boundary 
portrayed by the arbitrary trajectory $r=R(t)$, in 
a static bulk spacetime (\ref{staticm}). Hence all essential information 
of the bulk metric is transferred onto the boundary conditions 
and vice-versa. 

As a final remark, we note that the generalization to domain walls in
arbitrary dimension is straightforward, simply 
altering the $\mu/r^2$ to $\mu/r^{(n-3)}$ in (\ref{hrdef})
(see \cite{CW} for a different approach).

\section{Cosmological evolution equations}

In this section we derive the equations of motion for a general (i.e.\ not 
necessarily symmetric) wall bounding two regions of spacetime with
possibly differing cosmological constants. The reason for setting up
the problem in its full generality is that we will be able to use
the same set of equations to deal not only with non $Z_2$ symmetric
walls, such as in the Millenium model, but also with walls which
separate negative and zero cosmological constant spacetimes.
This latter case will of course also require consideration
of both class I and class II solutions for the $\Lambda=\kappa=0$ spacetime.
We begin with a wall bounding two class II spacetimes.

The starting point is the general solution (\ref{staticm}). 
To get a wall solution, we simply take a general boundary
\be
X^a = (t(\tau), R(\tau), \chi, \theta, \phi)
\ee
and compute the extrinsic curvatures on each side. 
Note that $\tau$ is the proper time with respect to an observer comoving
with the wall, i.e.\
\be\label{normcdn}
{\dot t}^2 h(r) - {\dot R}^2 h^{-1}(r) = 1
\ee
and is therefore well defined and  intrinsic to the wall, whereas 
$t$ is the coordinate time in the bulk, 
and need not agree with the coordinate time 
on the other side of the wall.
The only other subtlety
arises in the choice of sign for the outward normal
\be
n_a = \pm ( {\dot R}(\tau), -{\dot t}(\tau), {\bf 0})
\ee
This corresponds to the fact that 
we have a choice of which part of the spacetime we wish to keep (i.e.\
$r<R$ or $r>R$). This depends on the physical solution required, i.e.\
whether we have a positive or negative energy wall, and what sort of
asymptotic solution we require. In general, positive energy
walls match across two interior spacetimes and correspond to the lower ($-$)
choice of sign in the outward normal and vice versa. It is also possible
to have a wall matching an interior and exterior spacetime (for example the 
spherical collapsing wall, \cite{IS}, used by Gogberashvili \cite{G} in
an early version of the RS scenario), however, for
clarity in what follows {\it unless explicitly stated}, the formulae will 
be valid for +($-$) energy walls matching two interior (exterior) patches
of the Schwarzschild-ADS spacetime.

Computing the extrinsic curvature components on each side of the wall gives
\bml\label{iseqns}\bea
\E &=& \mp {3\over R} \left ( {\dot t}_+ h_+ + {\dot t}_- h_- \right ) 
\label{iseq1}\\
{2\over 3} \E - \T &=& \pm \left (
{{\ddot R} + {\half}h_+' \over {\dot t}_+ h_+} +
{{\ddot R} + {\half}h_-' \over {\dot t}_- h_-} \right)
\label{iseq2}
\eea\eml
where $t_\pm$ represent the coordinate times on each side of the wall,
${\dot t}_\pm h_\pm$ being given by (\ref{normcdn}) in each case.
Note that we can also express (\ref{iseq2})
in a different fashion as an energy conservation equation, 
\be\label{econs}
{D\E\over d\tau} + 3{{\dot R}\over R} (\E - \T) = 0
\ee

These represent the general equations of motion for an arbitrary wall
separating two regions of spacetime with possibly different cosmological
constants. Note that at this point \E\ and \T\ are
unconstrained physically, and are simply given by the above expressions
in terms of $R(\tau)$, which, other than representing a non-spacelike
hypersurface, is completely arbitrary. This is worth stressing -- that
we can take {\it any} such $R(\tau)$, and we will get a wall solution
with energy and tension determined by the embedding. Only when we
impose physical constraints on our energy and tension do we get constraints
on our possible hypersurface $R(\tau)$. Before exploring such constraints
however, we simply note that (\ref{iseqns}) can also be expressed
in terms of brane cosmological evolution equations (cf.\ Binetruy
et.\ al.\ \cite{BDEL}) which are
\bml\label{coseveqs}\bea
&&\left ({{\dot R}\over R} \right )^2 = {\E^2 \over 36}
+ {9(\Delta k^2)^2 \over 4 \E^2} - {\overline {k^2}} - {\kappa\over R^2} 
+ {{\bar \mu}\over R^4} - {9\Delta k^2 \Delta \mu \over 2 R^4\E^2}
+ {9 (\Delta \mu)^2 \over 4 \E^2 R^8}\label{friedbr}\\
&&{\ddot R} =  - (2\E - 3\T) \left ( {R\E\over36}+
{9\Delta k^2 \Delta \mu \over 2\E^3R^3} \right ) 
+ R \left ( {9(\Delta k^2)^2 (4\E - 3\T)\over 4\E^3} 
- {\overline {k^2}} \right ) - {{\bar \mu} \over R^3} 
-{27\T (\Delta\mu)^2 \over 4 \E^3 R^7}\label{raychbr}
\eea\eml
where we have now put $6k^2_\pm=-\Lambda_\pm$.
These are the completely general brane cosmological evolution equations
for a brane separating two `ADS' regions, 
and were first derived by Stoica et.\ al.\ \cite{STW}. 

Since we are particularly interested in the cosmological solutions for
branes in models with regions of vacuum spacetime, we must also 
consider the case of a planar wall bordering on a class I
spacetime which we may write as
\be
ds^2 = H(v) dr dv - r^2 d{\bf x}^2
\ee
This spacetime is actually flat \cite{taub}, as can be seen by the coordinate
transformation
\bml\label{starred}\bea
{\bf x}^* &=& r {\bf x} \\
t^* - z^* &=& r \\
t^* + z^* &=& \int H dv + r {\bf x}^2
\eea\eml
which gives the Minkowski metric in the starred coordinate system. 

Clearly to match with an ADS spacetime we need to set $r = R(\tau)$, and
compute the relevant extrinsic curvature quantities. For a wall bounding
a class II and class I spacetime, we find that the junction conditions
are now
\bml\bea
\E &=& \mp {3\over R} \left ( {\dot R} + {\dot t}_- h_- \right ) \\
{2\over 3} \E - \T &=& \pm \left (
{{\ddot R} \over {\dot R}} +
{{\ddot R} + {\half}h_-' \over {\dot t}_- h_-} \right)
\eea\eml
Therefore we now find a somewhat different set of cosmological equations 
for the evolution of the wall: 
\bml\label{cl1cl2eq}\bea
{{\dot R}\over R} &=& \pm \left ( {3k^2\over 2\E} - {\E\over 6} 
- {3\mu_-\over 2\E R^4} \right ) \\
{\ddot R} &=& {k^2R\over 2\E^2} (9k^2-\E^2) +
{(2\E - 3\T)R\over 36\E^3} (81k^4-\E^4) + {\mu_-\over2\E^3R^3}
\left [ 9k^2 (3\T-2\E) - \E^3 \right] - {27\T\mu_-^2\over4\E^3R^7}
\eea\eml
where we see that the `Friedman' equation is now linear in $\dot R$. 

Finally, if we are matching two class I spacetimes, the junction conditions
give the particularly simple relations
\bml\label{2class1eq}\bea
\E &=& \mp {6{\dot R}\over R} \\
\T &=& \mp \left ( {4{\dot R}\over R} + {2{\ddot R}\over{\dot R}} \right )
\eea\eml

In the next section we will compute cosmological trajectories for general 
walls, however, before doing so it is useful to verify that this approach 
is indeed valid by cross-checking it against a few simple known solutions. 
For example, a planar symmetric domain wall in pure Einstein gravity is 
known to have a particularly simple interpretation in terms of the matched 
interiors of two hyperboloids in Minkowski spacetime \cite{GWG}.
In the context of these equations, a $\Lambda=\kappa=0$ spacetime can
match class I or class II spacetimes. For the matching of two class
I spacetimes we must use (\ref{2class1eq}) with $\E=\T$. Clearly
this has the solution $R = e^{\E\tau/6}$, giving a de-Sitter like
induced metric in agreement with the standard domain wall metric
in wall-based coordinates (see e.g.\ \cite{IS,vil}). Inverting this
relation to find $R(v)$ in terms of the bulk coordinates gives
$R = -18/\E^2v$. Finally, inverting (\ref{starred}) gives the particularly
simple bulk equation of motion for the domain wall trajectory:
$t^{*2} - z^{*2} - {\bf x}^{*2} = - 36/\E^2$, i.e.\ a hyperboloid in
Minkowski spacetime. The `planar' domain wall is therefore a boundary between
the interior of two Minkowski hyperboloids as required.

\section{Cosmological wall solutions}

In this section we apply the general equations worked out in the 
previous section to find a variety of cosmological solutions. 
Let us begin by considering the particularly simple equation of state
of a domain wall, $\E=\T$, and find the most general trajectories.
First of all, note that (\ref{econs}) implies that \E\ is a constant,
therefore the brane-Friedmann equation, (\ref{friedbr}), can be written
as
\be\label{dwfried}
{R^4\over 4} \left ( {dR^2\over d\tau} \right )^2 =
aR^8-\kappa R^6 + b R^4 + c
\ee
where 
\bml\label{dwabc}\bea
a &=& {\E^2\over36} + {9(\Delta k^2)^2 \over 4 \E^2} - \overline{k^2}\\
b &=& \overline{\mu} - {9\Delta k^2 \Delta \mu \over 2 \E^2} \\
c &=& {9 (\Delta \mu)^2 \over 4 \E^2 }
\eea\eml
which can be integrated in general, giving $R^2$ in terms of elliptic functions,
although the expression is not particularly illuminating. If one of 
$\kappa,c$ is zero, the solutions are very simple to write down, but
for general $a,b,c,\kappa$, the qualitative behaviour of the solutions can be
deduced from (\ref{dwfried}). 
The constant `$a$' can be seen to measure the departure from criticality
of the domain wall -- i.e.\ that value which allows a static
planar domain wall solution in the absence of an ADM mass term. We will
therefore denote $a=0$ walls as critical, and $a>(<)\,0$ walls as
super-(sub-)critical.

Clearly, if there is a nonzero ADM mass
in the bulk there is always a solution which expands outward from an 
initial singularity, $R=0$, which is where the wall trajectory touches the
central singularity of the black hole. Whether or not this solution
expands indefinitely, or there is a final singularity depends on the roots
of (\ref{dwfried}). For $a\geq 0$, there are zero or two (possibly repeated)
roots for $R^2>0$. If $b>{9\over32a}$ for $\kappa = 1$, or $b>0$ otherwise, 
then there are no positive roots and the cosmology expands indefinitely. 
Otherwise there can be two separate roots, in which case there is a
cosmology with initial and final singularities, and a nonsingular cosmology
which contracts in to a minimum radius and re-expands. If there is a 
repeated root, $R_c^2$, then the cosmology asymptotes $R_c$ exponentially at
late times either as a contracting branch, or an expanding branch with an
initial singularity. If $a<0$ there is always one root and our universe starts
and ends its life on the black hole singularity. Finally, there is always
a static solution at any positive root of the quartic.

For planar domain walls, i.e.\ $\kappa=0$, we have the exact solution
\be\label{pdwsolns}
R^4 = \cases{ {1\over 4a} \left ( e^{4\sqrt{a} (\tau - \tau_0)} - 2b 
+ (b^2-4ac) e^{-4\sqrt{a} (\tau - \tau_0)} \right ) & $a>0$ \cr
{\sqrt{b^2-4ac}\over 2|a|} \sin \left \{ 4\sqrt{|a|} (\tau - \tau_0)\right \}
-{b\over 2|a|} & $a<0$ \cr
4b(\tau-\tau_0)^2 - {c\over b} & $a=0$\cr}
\ee
which clearly demonstrates the expansion outwards from an initial singularity,
and for $a>0$ shows the late time inflationary nature of the cosmology.

If $c=0$, i.e.\ the mass terms on either side of the wall are the
same, the exact solutions are
\be
\label{c=0}
R^2 = \cases{ {1\over 4a} \left ( e^{2\sqrt{a} (\tau - \tau_0)} + 2\kappa 
+ (\kappa^2-4ab) e^{-2\sqrt{a} (\tau - \tau_0)} \right ) & $a>0$ \cr
{\sqrt{\kappa^2-4ab}\over 2a} \sin \left \{ 2\sqrt{|a|} (\tau - \tau_0)\right \}
+{\kappa\over 2a} & $a<0$ \cr
-\kappa(\tau-\tau_0)^2 + {b\over \kappa} & $a=0$\cr}
\ee
where we have taken $\kappa\neq 0$: if $\kappa=0$ use (\ref{pdwsolns}).
A nice analysis of supercritical domain wall solutions and
their bulk embeddings was given by Khoury et.\ al.\ in \cite{KSW}.

For the case of a general wall we need some sort of equation of 
state $\T(\E)$, for which we will 
follow convention in setting
\be
\E = \E_0 + \rho \;\;\; ; \;\;\; \T = \E_0 - p
\ee
where $p = (\gamma-1)\rho$ would be a typical equation of state, although
we re-emphasize that for $\rho \simeq \E$ there is no reason to suppose
that such an equation of state will be accurate, or even possible.
The conservation of energy momentum (\ref{econs}) implies 
that $\rho =\rho_0 R^{-3\gamma}$ and hence 
$\E$ is no longer a constant. In order to deal with this additional difficulty
(in the non-$Z_2$-symmetric case) we keep only terms linear in $\rho$ 
which boils down to looking at wall trajectories
for  late times. For example in 
the case of radiation cosmology, $\gamma = 4/3$, 
the Friedmann equation 
(\ref{friedbr}) reduces to,
\be
\left ({{\dot R}\over R} \right )^2 = {\E_0^2\over36} + {9(\Delta k^2)^2 
\over 4 \E_0^2} - \overline{k^2}- {\kappa \over R^2}+ {1 \over R^4}\left(
\overline{\mu} - {9\Delta k^2 \Delta \mu \over 2 \E_0^2}+{\E_0\rho_0\over 18}
-{9\rho_0(\Delta k^2)^2 \over 2 \E_0^3}\right)+O(R^{-8})
\ee
The above equation has exactly the same solution as (\ref{c=0}) for 
$\kappa\neq 0$ where however 
the constants $a$ and $b$ are given by,
\be
a = {\E_0^2\over36} + {9(\Delta k^2)^2 \over 4 \E_0^2}- \overline{k^2}
\;\;\;\; ; \;\;\;\;
b = \overline{\mu} - {9\Delta k^2 \Delta \mu \over 2 \E_0^2}
+{\E_0\rho_0\over 18}
-{9\rho_0(\Delta k^2)^2 \over 2 \E_0^3}\;\;\;\; ; \;\;\;\;
\ee

Another important class of solutions are those with $Z_2$ symmetry, for 
which the cosmological Friedman equation has  the form of (\ref{dwfried})
with the differences `$\Delta$' of all quantities being zero. 
The Friedmann 
equation (\ref{friedbr}) becomes
\be
\label{z2fried}
\left ({{\dot R}\over R} \right )^2 = {\E_0^2\over36} - k^2
- {\kappa\over R^2} + {\E_0\rho_0\over 18 R^{3\gamma}} +{\mu\over R^4}
+ {\rho_0^2 \over 36 R^{6\gamma}}
\ee
where the above is now an exact expression.
For example, a radiation cosmology would have $\gamma = 4/3$, and the
equation of motion would be as (\ref{dwfried}), but with the constants
$a_r,b_r,c_r,$ now defined as
\be\label{z2abc}
a_r = {\E_0^2\over36} - k^2\;\;\;\; ; \;\;\;\;
b_r = \mu + {\E_0\rho_0\over 18} \;\;\;\; ; \;\;\;\;
c_r = {\rho_0^2\over 36} 
\ee
and so a generic planar solution would be given by (\ref{pdwsolns}) with these
new values for the constants $a,b$, and $c$.
We see now the interplay between the contribution of the cosmological matter
energy density $\rho_0$, which contributes to both $b$ and $c$, and the
``CFT'' contribution, $\mu$, which appears only in $b$.
This changes the way in which the initial singularity, $R=0$, is approached,
for example, Gubser's solution has $a = 0 = \rho_0$, and 
$R = \mu^{1\over4} \sqrt{2(\tau-\tau_0)}$, as opposed to 
$R = b^{1\over4} [4(\tau-\tau_0)^2 - c/b^2]^{1\over4}$.
A description of $Z_2$ symmetric cosmological solutions and 
their embeddings was given by Mukohyama et.\ al.\ \cite{MSM}.

Let us now turn to the less well-explored question of cosmological 
solutions in models with slices of vacuum and ADS spacetimes, such as the 
noncompact GRS model, or compact models such as those in \cite{KMPR2,LMW}.
Without loss of generality, we will take $k_+=0$, $k_-=k$, for which the 
general equations of motion are given by (\ref{coseveqs}) or 
(\ref{cl1cl2eq}). For simplicity, since we are unlikely to be 
interested in cosmologies on negative tension walls, let us 
consider the case where the boundary has the equation of
state of a domain wall, i.e.\ $\E=\T$, and let $\varepsilon = -(\E+3k)$ be
the departure from `criticality' of this domain wall ($\varepsilon>0$
being supercritical). 

We start by considering the matching of two class II spacetimes. Note that
for $\kappa = 0, -1$, in order to have the correct spacetime signature,
we require $\mu_+<0$, and we have a naked singularity in the full bulk 
solution. For $\kappa = 1$, $\mu_+>0$ gives the five-dimensional
Schwarzschild solution.

Substituting the values of $k_\pm$ gives for the constants in (\ref{dwfried})
\be
a = {\varepsilon^2 \over 36 \E^2} \left ( \varepsilon + 2\E\right )^2 \;\;;
\;\; b = \mu_+ + {\varepsilon(\varepsilon+2\E) \Delta \mu \over 2 \E^2} \;\;;
\;\; c = {9 (\Delta \mu)^2 \over 4 \E^2 }
\ee
Note that unlike the generic domain wall case, $a>0$ for
both sub and super-critical walls.

For $\kappa = 1$, $\mu_+>0$ and we again have solutions which expand 
out (or contract into) the singularity at $r=0$. Depending on the roots
of the quartic, there are solutions which expand indefinitely, recontract,
or indeed asymptote a constant radius. The critical domain wall always
recollapses. For $\kappa = -1$, $\mu_+<0$, and the qualitative
behaviour is similar, however the critical domain wall now expands indefinitely
if $27c>-4 \mu_+$. 

For the planar ($\kappa = 0$) walls, the exact solutions are
\be\label{grssolns}
R^4 = \cases{ {1\over 4a} \left ( e^{\pm4\sqrt{a} (\tau - \tau_0)} - 2b 
+ (b^2-4ac) e^{\mp4\sqrt{a} (\tau - \tau_0)} \right ) & $a>0$ \cr
4b(\tau-\tau_0)^2 - {c\over b} & $a=0$\cr}
\ee

\begin{figure}
\centerline{\epsfig{file=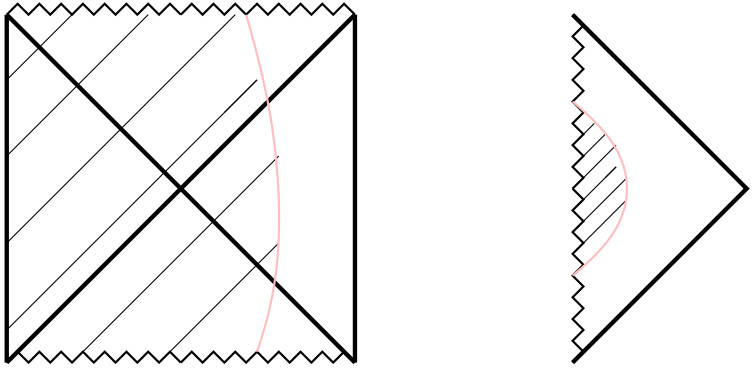,width=10cm}}
\vskip 5mm
\caption{A negative tension domain wall patching vacuum to Schwarzschild-ADS
spacetime. The vacuum bulk has a naked singularity.}
\label{cosgrs1}
\end{figure}

For the critical domain wall, $b = \mu_+ <0$, and we see that the
domain wall universe
has an initial and final singularity, for $\tau_0 - {\Delta\mu\over4k\mu_+}
< \tau < \tau_0 + {\Delta\mu\over4k\mu_+}$. Inverting (\ref{normcdn}) to
find the trajectory in terms of coordinate time in the vacuum spacetime to
the right of the wall shows that these singularities occur at finite coordinate
time, and the wall actually expands out of, and re-collapses into, the
central naked singularity in this spacetime. This corresponds to figure
\ref{cosgrs1},
and would appear to be
a rather undesirable spacetime from the five-dimensional point of view.

For the subcritical domain wall however, $b<0$ with $b^2-4ac>0$, hence
the cosmology has no initial singularity, but merely follows a modestly
corrected $\sinh$ trajectory. For the supercritical domain wall, there
are two possibilities, depending on the relative magnitudes of $\mu_\pm$. 
If $9k^2|\mu_+| \geq \mu_- (\varepsilon^2 + 6k\varepsilon)$, then the cosmology
is completely nonsingular as for the subcritical case, however, if
$9k^2|\mu_+|<\mu_- (\varepsilon^2 + 6k\varepsilon)$, then there is an initial
(or final) singularity with the universe inflating away (deflating towards)
it.

Finally, for a planar asymptotically vacuum spacetime, the other
possibility is matching to a class I spacetime for which we need
(\ref{cl1cl2eq}). Setting $\E=\T = -\varepsilon - 3k <0$ again we have
\be
{{\dot R}\over R} 
= {\varepsilon(\varepsilon+2\E)\over6\E} - {3\mu_-\over 2\E R^4}
\ee
which has the general solution
\be
R^4 = \cases { \exp \left \{ {2\varepsilon(\varepsilon+2\E)\over 3\E}
(\tau-\tau_0) \right \} + {9\mu_-\over \varepsilon(\varepsilon+2\E)}
& $\varepsilon \neq 0$ \cr
{2\mu_-\over k} (\tau-\tau_0) & $\varepsilon = 0 $ \cr}
\ee

A supercritical wall therefore has ${\dot R}>0$ and expands outward 
from an initial singularity (where it touches the Schwarzschild
singularity on the ($-$) side of the wall). A subcritical wall
however has ${\dot R}^>_< 0 $ for ${R^4}^<_> {9\mu_- \over \varepsilon(
\varepsilon + 2\E)}$. This shows that there are two solutions, one
of which has an inital singularity and expands outward to
$R_c^4 = {9\mu_- \over \varepsilon(\varepsilon + 2\E)}$, and
another which contracts to $R_c$. A critical domain wall has a power law
expansion outward from an initial singularity.

It is interesting to transform into the starred coordinate system 
to follow these wall trajectories in Minkowski spacetime:
\bml\label{msttraj}\bea
{\bf x}^* &=& R(\tau) {\bf x} \\
t^* - z^* &=& R(\tau) \\
t^* + z^* &=& \int {d\tau\over {\dot R}(\tau)} + R(\tau) {\bf x}^2
=F(R(\tau)) + R(\tau) {\bf x}^2
\eea\eml
For a critical domain wall $F(R)= 4 k^2 R^7 / 7\mu_-^2$, and for a 
noncritical domain wall bordering pure ADS, $F(R) = -1/4R$. We see
therefore that this latter case is the familiar hyperboloid in
Minkowski spacetime, although it is now the {\it interior} which is excised
for this negative tension wall.

\begin{figure}
\centerline{\epsfig{file=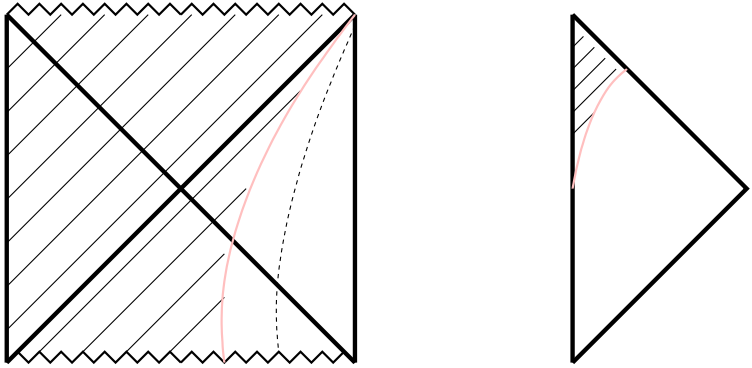,width=10cm}}
\vskip 5mm
\caption{A negative tension domain wall patching a class I (flat) vacuum 
spacetime to Schwarzschild-ADS. The dotted line indicates a possible 
additional positive energy cosmological domain wall.}
\label{cosgrs2}
\end{figure}

The critical domain wall satisfies
\be
{\bf x}^{*2} = t^{*2} - z^{*2} - {4k^2\over 7\mu_-^2} (t^* - z^*)^8
\ee
which is a deformed hyperboloid. At fixed time, the 3-sphere is now
squashed into an egg-shape in the $|{\bf x}^*|,z^*$ directions.
Again, it is the interior of this squashed S$^3$ which is excised.
On the ADS side, the wall expands outward from the initial black
hole singularity and at late times expands as $R \propto t_-^{1/3}$
(c.f.\ $R\propto t$ for Gubser's critical $Z_2$ symmetric solution).
The appropriate matching is shown in figure \ref{cosgrs2}.

Finally it is worth stressing that these trajectories are independent of
whether or not there is an additional positive tension wall present --
the existence of the wall trajectory is a local question. We can therefore
add in a positive domain wall in the usual way  by simply adding 
another boundary to the Schwarzschild ADS spacetime as shown by 
the dotted line in figure \ref{cosgrs2}. 

\section{Discussion}

Starting from the most general five-dimensional metric with homogeneous
and isotropic spatial 3-sections, we have shown that the most
general cosmological brane solution matches two 
Schwarzschild-ADS spacetimes, or a Schwarzschild-ADS spacetime with a 
class I or II vacuum spacetime. It is worth emphasizing that the 
three dimensional homogeneity and isotropy ensures that the 
bulk metric solution is invariant under two dimensional conformal
symmetries.  By inputting a boundary or wall we in principle 
break these conformal symmetries, however, it turns out that 
the junction conditions are such that only half are broken. This in turn
ensures that the only 
remaining degree of freedom is the wall's trajectory itself. Hence all the 
essential information of the wall's dynamics is contained on the wall 
trajectory itself (the boundary) the bulk being a specific static spacetime 
playing a `background' role.    

We then derived the generalised FRW equations for
a wall with general energy and tension, then applied these to a range of
cases, focusing on the previously unexplored case of a wall bordering
a vacuum region on one side. The interesting feature of these latter solutions
is that if two class II spacetimes are matched, the bulk can contain timelike
naked singularities on the vacuum side of the wall. For a more satisfactory
nonsingular solution, one must match the planar wall to a class I spacetime. 
Here the wall trajectories are reminiscent of the vacuum domain wall of
Vilenkin, Ipser and Sikivie, in that they are deformed hyperboloids, although
it is the interior of the hyperboloid which is excised for this negative
energy wall. 

If we now wish to construct a `cosmological' GRS solution, i.e.\ one which 
has a $Z_2$ symmetric positive energy central wall with matter residing
on it, and a negative energy outer wall, then we must combine one of the
$Z_2$ solutions of (\ref{z2fried}) with a class II/II or class II/I solution
for the outer wall. In each case, we must be careful to keep the ($-$) wall
to the right of the ($+$) wall, $R_-<R_+$. For example, if we take the
planar domain wall, the central domain wall trajectory, $R_+$, is given
by (\ref{pdwsolns}) with the constants $a,b,c$ in (\ref{z2abc}). Using the 
class II/I solutions of (\ref{grssolns}) for the ($-$) wall shows
that we cannot match arbitrary walls, since for example a subcritical
central wall would collide with a critical or supercritical outer
wall bringing our universe to an abrupt and catastrophic end.
Two generic critical walls are compatible however, with
\be
\cases{ R_+^4 = 4 \left ( \mu + {k\rho_0\over3}\right)
(\tau_+-\tau_{_0+})^2 - {\rho_0^2 \over 36\mu + 12k\rho_0} &\cr
R_-^4 = {2\mu\over k} (\tau_- - \tau_{_0-}) & \cr}
\ee
as is a critical central wall with a subcritical outer one.
A supercritical central wall can exist with a critical or subcritical 
outer wall, or a supercritical outer wall with $ 2\sqrt{\E_+^2-36k^2} > 
(9k^2-\E_-^2)/|\E_-|$; and a subcritical central wall can exist with
a subcritical outer wall provided $ {\sqrt{b_r^2 - 4a_rc_r} - b_r\over2|a_r|}
> {9\mu \over (9k^2 - \E_-^2)}$. 

For example we can match the critical Gubser solution,
$R_+^4 = 4\mu\tau_+^2$, with the critical class II/I solution
$R_-^4 = 2\mu \tau_-/k$. However, a pure critical planar radiation
cosmology, $R_+^4 = \rho_0(16k^2 \tau_+^2 -1)/12k$ cannot be matched to any
class I asymptotically flat spacetime across a ($-$) wall, as the only possible solution is an exponentially expanding or contracting one for a super/sub
critical wall respectively.

Therefore, while there are some constraints on the central wall
cosmologies, these are not overwhelmingly restrictive.

This leads us to the issue of the radion in these cosmological models. There
has been a great deal of debate about the radion in the context of cosmological
models (see e.g.\ \cite{COSRAD}) and in the case of quasi-localised gravity
(see e.g.\ \cite{Probs,GRSPRZ,CEHT}). We have not supposed any 
stabilization mechanism, \cite{stab},
for the extra dimension, but simply looked for free cosmological solutions.
Since the radion leads to anti-gravity in the original GRS model, it is
interesting to see how it manifests itself here.

The wavefunction of the radion is straightforward to find in linearized
gravity, \cite{CGR}, and it behaves as a scalar field living on the brane 
worldvolume which satisfies a massless equation of motion,
$\partial^2 f = 0$. In the context of a homogeneous and isotropic
cosmology, this has the simple solution $f = f_0t$, where $t$ is the coordinate
time on the brane, and we have chosen to have zero displacement for $t=0$.
Now consider a supercritical $Z_2$-symmetric positive energy planar domain
wall in pure ADS spacetime; this follows a trajectory
\be
R = \exp \{ \sqrt{\E^2 - 36k^2} (\tau - \tau_0)/6 \} = \left [
1 - {k^2 \sqrt{\E^2 - 36k^2}\over \E} (t-t_0) \right]^{-1}
\ee
in terms of either the brane proper time, $\tau$, or the local bulk 
coordinate time, $t$. We see therefore that for small $t$ and 
$\sqrt{\E^2 - 36k^2}$,
\be
R \simeq 1 + {k \sqrt{3k}\sqrt{\E - 6k}\over 3} (t - t_0)
\ee
we can therefore identify $f_0 = k^{3/2}\sqrt{\E-6k}\sqrt{3}$, i.e.\
$\delta \E = \E-6k = 3f_0^2 k^{-3/2}$, which has the correct dependence
on $f_0$ for a standard energy momentum tensor for the radion.
(Note we cannot have a subcritical planar domain wall.) This means that
a negative energy wall requires $\E < -6k$, and nominally a negative
radion energy on that wall.

Obviously these solutions do not see any evidence of the fifth dimension
``opening up'' \cite{GRS}, or any instability due to the radion, however,
this is because of the high degree of symmetry of the set-up. If we were
to break isotropy or homogeneity, the integrable nature of the system would
likely be destroyed, and the simplicity of the description of the wall as
a trajectory in bulk Schwarzschild-ADS would disappear. Destruction of these
symmetries would lead to a brane-bulk interaction (as evidenced by the 
extreme case of the  as yet undiscovered black hole on the brane
solution -- see \cite{BH} for discussions on this topic) through which the
more familiar effects of the radion would be recovered.

Of course it is tempting to enquire whether the so-called ``missing mode''
of free motion of a wall can be identified in this set-up.
For a non-gravitating defect, the effective equation of motion is given
by the relation $\overline{K}=0$, i.e.\ the wall (in this case) is a
minimal hypersurface. An equivalent perturbative description would state
that the displacement of the wall satisfies a massless wave equation on
the wall. There have been claims \cite{II} in the literature
that when gravity is included this `free motion' disappears, as would
appear to agree with the rejection of the free `radion' solution
\cite{CGR}
\be
\delta g_{\mu\nu} = {(r^{-2} - r^2)\over 4k} {\overline{K}}_{\mu\nu}
\ee
in the usual Randall-Sundrum perturbation analysis. However, in this
simple picture, each wall has its own `radion' corresponding to motion
through the ambient spacetime. Indeed, a perturbative analysis of the 
thick four-dimensional domain wall \cite{BCG} indicates no such
disappearance of this motion with the coupling to gravity. What
therefore can be going on? A clue perhaps lies in the effective
equation of state for a freely moving wall. As Carter and Vilenkin
\cite{CV}, argued for the cosmic string, a freely moving defect
will have an altered effective equation of state. The effective
energy per unit area will increase, and the tension will decrease.
Generalizing the Carter-Vilenkin formula for the domain wall
gives the effective equation of state
\be
\E\T^3 = \E_0^4
\ee
for a 3-brane domain wall. Clearly for \E\ close to $\E_0$, this gives
an equation of state of a radiation cosmology on the background domain
wall. Now we see how a freely moving wall might no longer be a small
perturbation of a `straight' one: the averaged equation of state of
the freely moving wall causes the wall to follow a centre of mass trajectory 
associated with a radiation universe, which is not, at late times
(when the radiation universe approximation is particularly good)
a small perturbation of the unmoving wall. Of course, a freely moving
wall will not be isotropic and homogeneous except at the very large 
scale, and it is also possible that there are brane-bulk interactions
which further complicate the issue.

Finally, the problem of brane cosmological perturbation theory is
very important to understand if we wish to do real cosmology. The
setting up of the formalism for perturbation theory on the brane 
is already underway \cite{KIS,PTH}; by considering the 
cosmological domain wall in terms of its global spacetime structure, 
the role of the bulk, its interactions with matter on the brane, and
the interpretation of some of the gauge invariant variables in terms
of bulk physical quantities will be more clearly elucidated.


\section*{Acknowledgements}

It is a great pleasure to thank Roberto Emparan, David Fairlie, Jihad Mourad,
Simon Ross, Valery Rubakov, and Douglas Smith for helpful and enlightening 
discussions. 

C.C.\ was supported by PPARC, and R.G.\ by the Royal Society.


\begin{references}
\bibitem{RS} L.Randall and R.Sundrum, \prl 83 3370 1999. [hep-ph/9905221]\\
L.Randall and R.Sundrum, \prl 83 4690 1999. [hep-th/9906064]
\bibitem{RSA} V.A.Rubakov and M.E.Shaposhnikov, \plb 125 136 1983.\\
K.Akama, in {\it Gauge Theory and Gravitation.  Proceedings of the
International Symposium, Nara, Japan, 1982}, eds. K.Kikkawa, N.Nakanishi and
H.Nariai (Springer--Verlag, 1983).
\bibitem{EXO} M.Visser, \plb 159 22 1985. [hep-th/9910093].\\
E.J.Squires, \plb 167 286 1986.
\bibitem{HCD} A.G.Cohen and D.B.Kaplan, \plb 470 52 1999. [hep-th/9910132]\\
R.Gregory, \prl 84 2564 2000. [hep-th/9911015]\\
T.Gherghetta and M.Shaposhnikov, \prl 85 240 2000. [hep-th/0004014]\\
M.Gogberashvili and P.Midodashvili, {\it Brane universe in six dimensions},
hep-ph/0005298.\\
I.Olasagasti and A.Vilenkin, {\it Gravity of higher-dimensional
global defects}, hep-th/0003300.
\bibitem{HW} P.Horava and E.Witten, \npb 475 94 1996. [hep-th/9603142]
\bibitem{LOSW} A.Lukas, B.A.Ovrut, K.S.Stelle and D.Waldram, \prd 59 086001
1999. [hep-th/9803235]
\bibitem{LOW} A.Lukas, B.A.Ovrut and D.Waldram, \prd 60 086001 1999.
[hep-th/9806022]
\bibitem{HR} H.Reall, \prd 59 103506 1999. [hep-th/9809195]
\bibitem{GT} J.Garriga and T.Tanaka, \prl 84 2778 2000. [hep-th/9911055]
\bibitem{GKR} S.Giddings, E.Katz and L.Randall, JHEP 0003:023 (2000).
[hep-th/0002091]
\bibitem{BDL} P.Binetruy, C.Deffayet and D.Langlois,
\npb 565 269 2000. [hep-th/9905012]
\bibitem{BDEL} P.Binetruy, C.Deffayet, U.Ellwanger and D.Langlois, 
\plb 477 285 2000.  [hep-th/9910219]
\bibitem{COS} N.Kaloper, \prd 60 123506 1999. [hep-th/9905210]\\
D.J.Chung and K.Freese, \prd 61 023511 2000. [hep-ph/9906542]\\
C.Csaki, M.Graesser, C.Kolda and J.Terning, \plb 462 34 1999.
[hep-ph/9906513]\\
J.Cline, C.Grojean and G.Servant, \prl 83 4245 1999. [hep-ph/9906523]
\bibitem{SMS} T.Shiromizu, K.-I.Maeda and M.Sasaki, {\it The Einstein
equations on the 3-brane world}, gr-qc/9910076.
\bibitem{M} R.Maartens, {\it Cosmological dynamics on the brane}, 
hep-th/0004166.
\bibitem{Ida} D.Ida, {\it Brane world cosmology}, gr-qc/9912002.
\bibitem{MSM} S.Mukohyama, T.Shiromizu and K.-I.Maeda, \prd 62 024028 2000.
[hep-th/9912287]
\bibitem{PK} P.Kraus, JHEP 9912:011 (1999). [hep-th/9910149]
\bibitem{CRP} H.A.Chamblin and H.Reall, \npb 562 133 1999. [hep-th/9903225]\\
H.A.Chamblin, M.J.Perry and H.Reall, JHEP 9909:014 (1999). [hep-th/9908047]
\bibitem{Gub} S.Gubser, {\it ADS/CFT and gravity}, hep-th/9912001. \\
S.W.Hawking, T.Hertog and H.S.Reall, \prd 62 043501 2000. [hep-th/0003052]\\
L.Anchordoqui, C.Nunez and K.Olsen, {\it Quantum cosmology and AdS/CFT},
hep-th/0007064.
\bibitem{LR} J.Lykken and L.Randall, JHEP 0006:014 (2000). [hep-th/9908076]
\bibitem{IS} J. Ipser and P. Sikivie, \prd 30 712 1984.
\bibitem{KMPR} I.I.Kogan, S.Mouslopoulos, A.Papazoglou and G.G.Ross,
{\it A three three-brane universe: new phenomenology for the new millenium?},
hep-ph/9912552.
\bibitem{GRS} R.Gregory, V.A.Rubakov and S.Sibiryakov, \prl 84 5928-5931 2000.
[hep-th/0002072]
\bibitem{Probs} C.Csaki, J.Erlich and T.J.Hollowood, \prl 84 5932 2000.
[hep-th/0002161]\\
G.Dvali, G.Gabadadze and M.Porrati, \plb 484 112 2000. [hep-th/0002190]\\
C.Csaki, J.Erlich and T.J.Hollowood, \plb 481 107 2000. [hep-th/0003020]
\bibitem{GRSPRZ} R.Gregory, V.A.Rubakov and S.Sibiryakov, {\it Gravity 
and antigravity in a brane world with metastable gravitons}, hep-th/0003045.\\
L.Pilo, R.Rattazzi and A.Zaffaroni, {\it The fate of the radion in models
with metastable graviton}, hep-th/0004028.
\bibitem{CEHT} C.Csaki, J.Erlich, T.J.Hollowood and J.Terning,
{\it Holographic RG and cosmology in theories with quasilocalized gravity},
hep-th/0003076.
\bibitem{KMPR2} I.I.Kogan, S.Mouslopoulos, A.Papazoglou and G.G.Ross,
{\it Multi-brane worlds and modification of gravity at large scales},
hep-th/0006030.
\bibitem{COSRAD} P.Kanti, I.I.Kogan, K.A.Olive and M.Pospelov,
\plb 468 31 1999. [hep-ph/9909481]\\
E.Flanagan, S.H.Tye and I.Wasserman, {\it A cosmology of the brane world},
hep-ph/9909373.\\
E.Flanagan, S.H.Tye and I.Wasserman, {\it Cosmological expansion in the
Randall-Sundrum brane world scenario}, hep-ph/9910498.\\
C.Csaki, M.Graesser, L.Randall and J.Terning, {\it Cosmology of brane
models with radion stabilization}, hep-ph/9911406.\\
H.Collins and B.Holdom, {\it Brane cosmologies without orbifolds}, 
hep-ph/0003173.
C.Barcelo and M.Visser, \plb 482 183 2000. [hep-th/0004056]
\bibitem{taub} A.H.Taub, \am 53 472 1951.
\bibitem{Israel} W.Israel,  \ncim 44B 1 1966.
\bibitem{CW} M.Cvetic and J.Wang, \prd 61 124020 (2000). [hep-th/9912187]
\bibitem{G} M.Gogberashvili, \epl 49 396 2000. [hep-ph/9812365]
\bibitem{STW} H.Stoica, S.H.Tye and I.Wasserman, \plb 482 205 2000. 
[hep-th/0004126]
\bibitem{GWG} G.W.Gibbons, \npb 394 3 1993.\\
M.Cvetic, S.Griffies and H.Soleng, \prl 71 670 1993. [hep-th/9212020]\\
M.Cvetic, S.Griffies and H.Soleng, \prd 48 2613 1993. [gr-qc/9306005]
\bibitem{vil} A.Vilenkin, \plb 133 177 1983.
\bibitem{KSW} J.Khoury, P.Steinhardt and D.Waldram, {\it Inflationary
solutions in the brane world and their geometrical interpretation},
hep-th/0006069.
\bibitem{LMW} J.Lykken, R.C.Myers and J.Wang, {\it Gravity in a box},
hep-th/0006191.
\bibitem{stab} W.Goldberger and M.Wise, \prl 83 4922 1999. [hep-ph/9907447]\\
D.Choudhury, D.P.Jatkar, U.Mahanta and S.Sur, {\it On
the stability of the three 3-brane model}, hep-ph/0004233.\\
J.E.Kim and B.Kyae, hep-th/0005139.
\bibitem{CGR} C.Charmousis, R.Gregory and V.A.Rubakov, {\it Wave function of 
the radion in a brane world}, hep-th/9912160.
\bibitem{BH} A.Chamblin, S.W.Hawking and H.S.Reall, \prd 61 065007 2000.
[hep-th/9909205]\\
R.Emparan, G.T.Horowitz and R.C.Myers, JHEP 0001:007 (2000).
[hep-th/9911043]\\
R.Emparan, G.T.Horowitz and R.C.Myers, JHEP 0001:021 (2000). [hep-th/9912135]\\
A.Chamblin, C.Csaki, J.Erlich and T.J.Hollowood, \prd 62 044012 2000.
[hep-th/0002076]\\
R.Gregory, {\it Black string instabilities in ADS space}, hep-th/0004101.
\bibitem{II} A.Ishibashi and H.Ishihara, \prd 56 3446 1997. [gr-qc/9704058]\\
A.Ishibashi and H.Ishihara, \prd 60 124016 1999. [gr-qc/9802036]
\bibitem{BCG} F.Bonjour, C.Charmousis and R.Gregory, {\it The dynamics of
curved gravitating walls}, gr-qc/0002063.
\bibitem{CV} A.Vilenkin, \prd 41 3038 1990.\\
B.Carter, \prd 41 3869 1990.
\bibitem{KIS} S.Mukohyama, {\it Gauge invariant gravitational perturbations
of maximally symmetric spacetimes}, hep-th/0004067.\\
H.Kodama, A.Ishibashi and O.Sato, {\it Brane world cosmology --
gauge invariant formalism for perturbation}, hep-th/0004160.
\bibitem{PTH} D.Langlois, {\it Brane cosmological perturbations}, 
hep-th/0005025.\\
C.van de Bruck, M.Dorca, R.H.Brandenberger and A.Lukas, {\it Cosmological 
perturbations in brane world theories: formalism}, hep-th/0005032.\\
K.Koyama and J.Soda, {\it Evolution of cosmological perturbations in the
brane world}, hep-th/0005239.

\end{references}
\end{document}